\documentclass[showpacs,10pt,twocolumn,prb]{revtex4-1}

\usepackage{amsmath}
\usepackage{amssymb}
\usepackage{graphicx}
\usepackage{amssymb}
\usepackage{graphics}
\usepackage{epsfig}
\usepackage{color}

\begin{document}

\title{Critical current density and vortex pinning in tetragonal FeS$_{1-x}$Se$_{x}$ ($x=0,0.06$)}
\author{Aifeng Wang,$^{1}$ Lijun Wu,$^{1}$ V. N. Ivanovski,$^{2}$ J. B. Warren,$^{3}$ Jianjun Tian,$^{1,4}$ Yimei Zhu$^{1}$ and C. Petrovic$^{1}$}
\affiliation{$^{1}$Condensed Matter Physics and Materials Science Department, Brookhaven National Laboratory, Upton, New York 11973, USA\\
$^{2}$Institute of Nuclear Sciences Vinca, University of Belgrade, Belgrade 11001, Serbia\\
$^{3}$Instrument Division, Brookhaven National Laboratory, Upton, New York 11973, USA\\
$^{4}$School of Physics and Electronics, Henan University, Kaifeng 475004, China}
\date{\today}

\begin{abstract}
We report critical current density ($J_c$) in tetragonal FeS single crystals, similar to iron based superconductors with much higher superconducting critical temperatures ($T_{c}$'s). The $J_c$ is enhanced 3 times by 6\% Se doping. We observe scaling of the normalized vortex pinning force as a function of reduced field at all temperatures. Vortex pinning in FeS and FeS$_{0.94}$Se$_{0.06}$ shows contribution of core-normal surface-like pinning. Reduced temperature dependence of $J_c$ indicates that dominant interaction of vortex cores and pinning centers is via scattering of charge carriers with reduced mean free path ($\delta$$l$), in contrast to K$_x$Fe$_{2-y}$Se$_2$ where spatial variations in $T_{c}$ ($\delta$$T_{c}$) prevails.

\end{abstract}
\pacs{74.70.Xa,74.70.Ad,75.50.Lk,74.72.Cj}
\maketitle

\section{INTRODUCTION}

Fe-based superconductors have been attracting considerable attention since their discovery in 2008.\cite{Kamihara} Due to rich structural variety and signatures of high-temperature superconductivity similar or above iron arsenides, iron chalcogenide materials with Fe-Ch (Ch=S,Se,Te) building blocks  are of particular interest.\cite{WangQY,HeS,GeJF,ShiogaiJ} Recently, superconductivity below 5 K is found in tetragonal FeS synthesized by the hydrothermal reaction.\cite{LaiXF} The superconducting state is multiband with nodal gap and large upper critical field anisotropy.\cite{Borg,LinH,XingJ,YingTP} Local probe $\mu$SR measurements indicate two $s$-wave gaps but also a disordered impurity magnetism with small moment that microscopically coexists with bulk superconductivity below superconducting transition temperature.\cite{Holenstein} This is similar to FeSe at high pressures albeit with weaker coupling and larger coherence length.\cite{Khasanov1,Khasanov2}

Binary iron chalcogenides show potential for high field applications.\cite{SiW,SunY,LeoA,JungSG} Since FeCh tetrahedra could be incorporated in different superconducting materials, it is of interest to study critical currents and vortex pinning mechanism in tetragonal FeS.\cite{HosonoS,ForondaFR,LuXF} Moreover, vortex pinning and dynamics is strongly related to coherence length and superconducting pairing mechanism.

Here we report critical current density and the vortex pinning mechanism in FeS and and FeS$_{0.94}$Se$_{0.06}$. In contrast to the point defect pinning in Ba$_{0.6}$K$_{0.4}$Fe$_2$As$_2$ and K$_x$Fe$_{2-y}$Se$_2$,\cite{YangH,LeiHC2,LeiHC1} the scattering of charge carriers with reduced mean free path $l$ ($\delta$$l$ pinning) is important in vortex interaction with pinning centers.

\section{EXPERIMENTAL DETAILS}

FeS and FeS$_{0.94}$Se$_{0.06}$ single crystals were synthesized by de-intercalation of potassium from K$_x$Fe$_{2-y}$(Se,S)$_2$ single crystals, using the hydrothermal reaction method.\cite{LaiXF,LeiHC1} First, 8 mmol Fe powder, 5 mmol Na$_2$S$\cdot$9H$_2$O, 5 mmol NaOH, and 10 ml deionized water were mixed together and put into 25 ml Teflon-lined steel autoclave. After that, $\sim$0.2g K$_x$Fe$_{2-y}$S$_2$ and K$_x$Fe$_{2-y}$S$_{1.6}$Se$_{0.4}$ single crystals were added. The autoclave is tightly sealed and annealed at 120 $^{\circ}$C for three days. Silver colored FeS single crystals were obtained by washing the powder by de-ionized water and alcohol. Finally, FeS single crystals were obtained by drying in the vacuum overnight. X-ray diffraction (XRD) data were taken with Cu K$_{\alpha}$ ($\lambda=0.15418$ nm) radiation of Rigaku Miniflex powder diffractometer. The element analysis was performed using an energy-dispersive x-ray spectroscopy (EDX) in a JEOL LSM-6500 scanning electron microscope. High-resolution TEM imaging and electron diffraction were performed using the double aberration-corrected JEOL-ARM200CF microscope with a cold-field emission gun and operated at 200 kV. M\"{o}ssbauer spectrum was performed in a transmission geometry with $^{57}$Co(Rh) source at the room temperature. Single crystals are aligned on the sample holder plane with some overlap but without stack overflow. The spectrum has been examined by WinNormos software.\cite{Brand} Calibration of the spectrum was performed by laser and isomer shifts were given with respect to $\alpha $-Fe. Magnetization measurements on rectangular bar samples were performed in a Quantum Design Magnetic Property Measurement System (MPMS-XL5).

\section{RESULTS AND DISCUSSIONS}

\begin{figure}
\centerline{\includegraphics[scale=0.4]{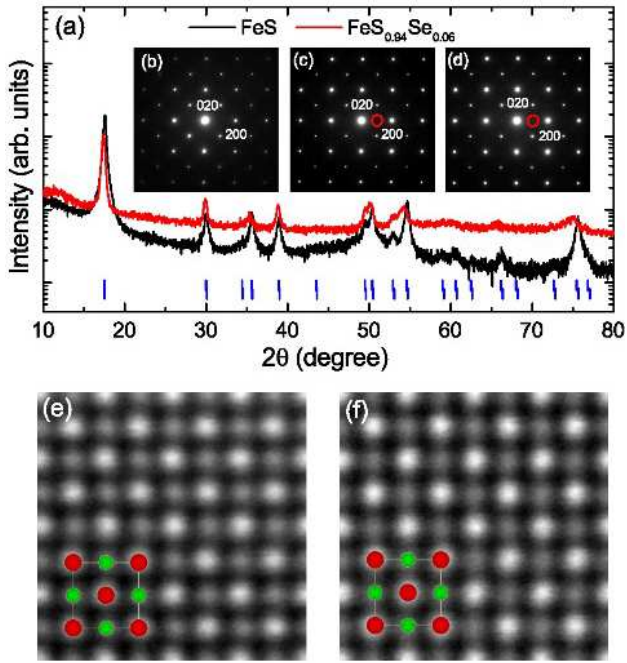}}
\caption{(Color online). (a) Powder x-ray diffraction pattern of tetragonal FeS (bottom) and FeS$_{0.94}$Se$_{0.06}$ (top). Vertical ticks mark reflections of the \textit{P4/nmm} space group. Electron diffraction pattern for FeS (b), and FeS$_{0.94}$Se$_{0.06}$ (c) and (d). High angle annular dark field scanning transmission electron microscopy (HAADF-STEM) image viewed along [001] direction of FeS (e) and FeS$_{0.94}$Se$_{0.06}$ (f) single crystal. [001] atomic projection of FeS is embedded in (b) with red and green spheres representing Fe and S/Se, respectively. The reflection condition in (b), (c) and (d) is consistent with P4/nmm space group. While the spots with h+k=odd are extinct in FeS, they are more [in (d)] or less [in (c)] observed here, indicating possible ordering of Se.}
\label{magnetism}
\end{figure}

Figure 1 (a) shows powder X ray diffraction pattern of FeS and FeS$_{0.94}$Se$_{0.06}$. The lattice parameters of FeS$_{0.94}$Se$_{0.06}$ are a=0.3682(2) nm and c=0.5063(3) nm, suggesting Se substitution on S atomic site in FeS [a=0.3673(2) nm, c=0.5028(2) nm]. High-resolution TEM imaging is consistent with the \textit{P4/nmm} unit cell and indicates possible ordering of Se atoms.

\begin{figure}
\centerline{\includegraphics[scale=0.3]{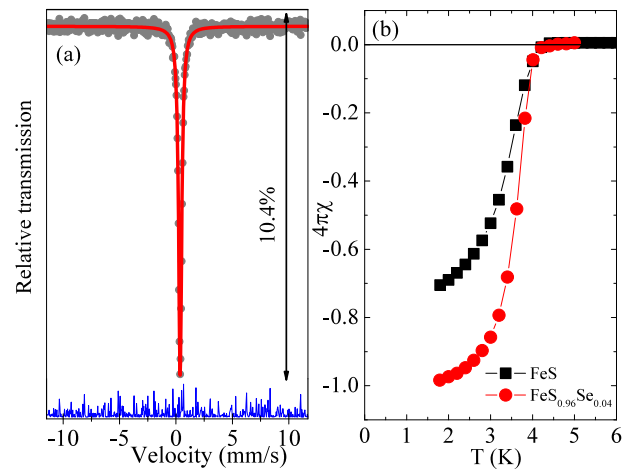}}
\caption{(Color online). (a) M\"{o}ssbauer spectrum at 294 K of the tetragonal FeS. The observed data are presented by the gray solid circles, fit is given by the red solid line, and difference is shown by the blue solid line. Vertical arrow denotes relative position of the experimental point with respect to the background. (b) Superconducting transition of FeS and FeS$_{0.94}$Se$_{0.06}$ measured by magnetic susceptibility in magnetic field 10 Oe.}
\label{magnetism}
\end{figure}

FeS M\"{o}ssbauer fit at the room temperature shows a singlet line [Fig.2(a)] and the absence of long range magnetic order. The isomer shift is $\delta$ = 0.373(1) mm/s whereas the Lorentz line width is 0.335(3) mm/s, in agreement with the previous measurements.\cite{MulletM,VaughanDJ,BertautEF}  Since FeS$_{4}$  tetrahedra are nearly ideal, one would expect axial symmetry of the electric field gradient (\textit{EFG}) and small values of the largest component of its diagonalized tensor $V_{zz}$. The linewidth is somewhat enhanced and is likely the consequence of small quadrupole splitting. If the Lorentz singlet would be split into two lines, their centroids would have been 0.06 mm/s apart, which is the measure of quadrupole splitting ($\Delta$). The measured isomer shift is consistent with Fe$^{2+}$, in agreement with X-ray absorbtion and photoemission spectroscopy studies.\cite{KwonK} There is very mild discrepancy of M\"{o}ssbauer theoretical curve when compared to observed values near 0.2 mm/s, most likely due to texture effects and small deviations of incident $\gamma$ rays from the c-axis of the crystal. The point defect corrections to M\"{o}ssbauer fitting curve are negligible. Fig. 2(b) presents the zero-field-cooling (ZFC) magnetic susceptibility taken at 10 Oe applied perpendicular to the $c$ axis for FeS and FeS$_{0.94}$Se$_{0.06}$ single crystals. Superconducting transition temperature Tc = 4.4 K (onset of diamagnetic signal) is observed in FeS, consistent with previous report.\cite{LaiXF} There is almost no change of $T_c$ in FeS$_{0.94}$Se$_{0.06}$. Both samples exhibit bulk superconductivity with somewhat enhanced diamagnetic shielding with Se substitution.

\begin{figure}
\centerline{\includegraphics[scale=0.44]{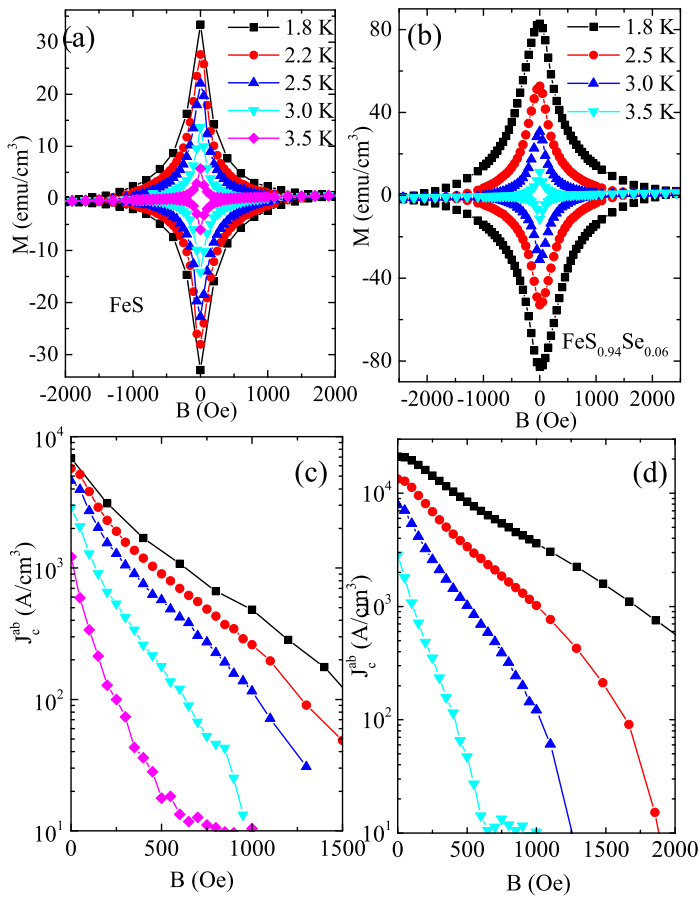}}
\caption{(Color online). Magnetization hysteresis loops at various temperatures for FeS (a), and FeSe$_{0.06}$ (b), respectively. Magnetic field was applied parallel to $c$ axis. Magnetic field dependence of the in-plane critical current density $J_c^{ab}$  for FeS (c), and FeSe$_{0.06}$ (d), respectively.}
\label{magnetism}
\end{figure}

Figures 3(a) and 3(b) show the magnetization hysteresis loops (MHL) for FeS and FeS$_{0.94}$Se$_{0.06}$, respectively. Both MHLs show symmetric field dependence and absence of paramagnetic background, suggestive of dominant bulk pinning.\cite{LinH} No fishtail effect is observed in both samples. The critical current density $J_c$ can be calculated from MHL using the Bean model.\cite{Bean} When the field is applied along $c$ axis, the in-plane critical current density $J_c(\mu_0H)$ is given by\cite{Bean, Gyorgy} \[{J}_{c} = \frac{20\Delta M(\mu_0H)}{a(1-a/3b)}\] where $\Delta M(\mu_0H)$ is the width of magnetization loop at specific applied field value and is measured in emu/cm$^{3}$. The $a$ and $b$ are the width and length of the sample ($a$$\leq$$b$) and measured in cm. The $J_c$ used in the formula is the unrelaxed critical current density. Practically measured critical current density, however, is the $J_s$ (relaxed value). Because magnetization relaxation is not very strong in iron-based superconductors, the $J_s$$\approx$$J_c$.\cite{ShenB} The paramagnetic (linear) $M(\mu_{0}H)$ background has no effect on the calculation of $\Delta$M($\mu_{0}$H) and consequently crucial currents due to its subtraction. The inclusion of ferromagnetic impurities\cite{Holenstein} is unlikely due to highly symmetric M(H) loops (Fig. 3). Therefore we attribute somewhat reduced volume fraction in pure tetragonal FeS to the presence of the unreacted paramagnetic hydrothermal solvent on the surface of our crystal, similar to what has been observed before.\cite{LaiXF}

Figs. 3(c) and 3(d) show the field dependence of $J_s$. The calculated $J_c$ at 1.8 K and 0 T reach 6.9 $\times$ 10$^3$ A/cm$^2$ and 2.1 $\times$ 10$^4$ A/cm$^2$, i.e. $J_c$ increases about 3 times for small Se substitution. It is interesting to note that the critical current densities of FeS and FeS$_{0.94}$Se$_{0.06}$ are comparable to that of K$_x$Fe$_{2-y}$Se$_2$ and FeSe which feature much higher superconducting transition temperatures (32 K and 8 K, respectively).\cite{GaoZS,LeiHC2,LeiHC3,Hafiez} However, $J_c$ of FeS and FeS$_{0.94}$Se$_{0.06}$ are lower when compared to iron pnictide  superconductors where typical critical current densities are above 10$^5$ A/cm$^2$ at 2 K.\cite{YangH}

\begin{figure}
\centerline{\includegraphics[scale=0.44]{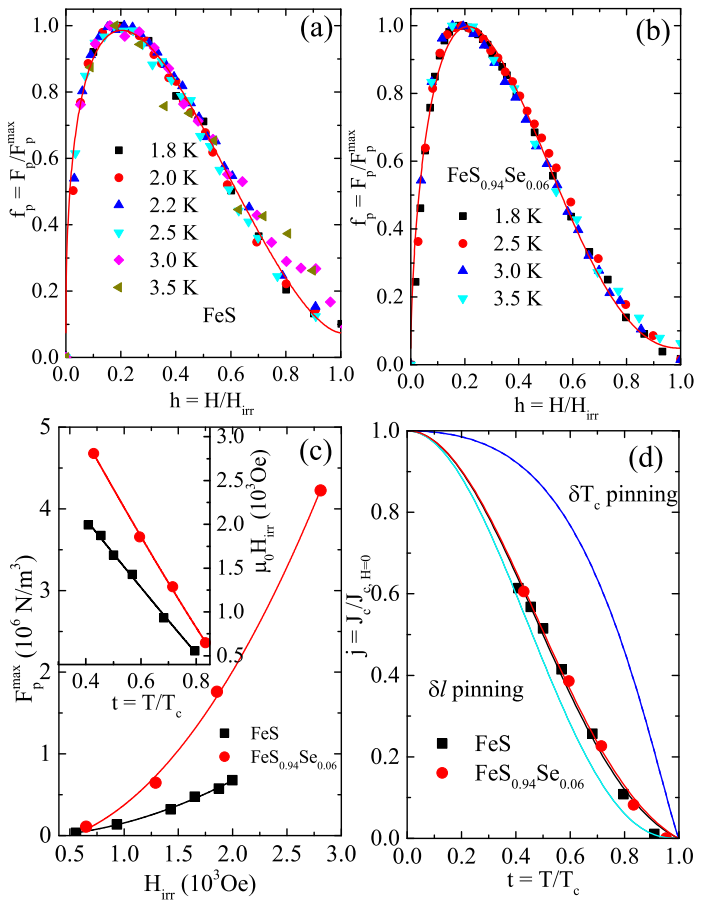}}
\caption{(Color online). Normalized flux pinning force as a function of the reduced field  for FeS (a), and FeSe$_{0.06}$ (b), respectively. Solid lines are the  fitting curves using $f_p$ = $Ah^p(1-h)^q$. (c) Maximum pinning force ($J_p^{max}$) vs field curves for FeS and FeS$_{0.94}$Se$_{0.06}$. The inset shows reduced temperature dependence of irreversibility field. (d) Reduced field dependence of normalized critical current at zero field for  FeS and FeS$_{0.94}$Se$_{0.06}$, blue and cyan solid lines correspond to theoretical data for $\delta$T$_c$ pinning and $\delta$$l$ pinning.}
\label{magnetism}
\end{figure}

Pinning force ($F_p = \mu_0HJ_c$) can provide useful insight into vortex dynamics. There is a peak in the pinning force density as a function of the reduced magnetic field for all hard high-field superconductors.\cite{Kramer} According to Dew-Hughes model,\cite{Hughes}, normalized vortex pinning force $f_p$ = $F_p/F_p^{max}$ should be proportional to $h^p(1-h)^q$, where $h = H/H_{irr}$ is normalized field, and $H_{irr}$ is irreversibility field obtained by an extrapolation of  $J_c(T, \mu_0H)$ to zero. Parameters $p$ and $q$ are determined by the pinning mechanism. As shown in Figs. 4(a) and 4(b), the curves of $f(h)$ at different temperatures overlap well with each other, indicating that the same pinning mechanism dominate at the temperature range we study. Fitting with scaling law $h^p(1-h)^q$ gives $p$ = 0.42, $q$ = 1.65, and $h_{max}^{fit}$ =0.21 for FeS and $p$ = 0.63, $q$ = 2.34, and $h_{max}^{fit}$ =0.21 for FeS$_{0.94}$Se$_{0.06}$, roughly consistent with theoretical $p$ = 0.5, $q$ = 2, $h_{max}^{fit}$ = 0.2 for core normal surface-like pinning. Core normal surface-like pinning describes the pinning center from the microstructure and geometry aspect. The free energy of the flux lines in the pinning centers is different from that in the superconducting matrix and the pinning center is normal whereas the geometry of the pinning centers is two dimensional. Weak and widely spaced pins induce a small peak in $f(h)$ at high $h$, while strong closely spaced pins produce a large peak at low $h$.\cite{Kramer} Similar $h$ indicates the strength and spacing of the pins is similar between two samples.

Figure 4(c) presents irreversibility field ($\mu_0H_{irr}$) dependence of $F_p^{max}$. Both the pinning force and $\mu_0H_{irr}$ are enhanced by Se doping. The curves can be scaled by $F_p^{max} \propto \mu_0H_{irr}^{\alpha}$ with $\alpha$ = 2.0 for FeS and 2.1 for FeS$_{0.94}$Se$_{0.06}$, close to theoretical prediction ($\alpha$ = 2).\cite{Hughes} Since the tetragonal FeS and FeS$_{0.94}$Se$_{0.06}$ were synthetized by de-intercalation of potassium using hydrothermal method and are cleaved along the $c$ axis much easier than other iron superconductors, it is likely that weakly connected surfaces are important in the flux pinning as opposed to K$_{x}$Fe$_{2-y}$Se$_{2}$, FeSe$_{0.5}$Te$_{0.5}$ thin flim, and Ba$_{0.6}$K$_{0.4}$Fe$_2$As$_2$  where point-like pinning prevails.\cite{LeiHC2,YangH,YuanPS}

In addition, as shown in the inset of Fig. 4(c), the reduced temperature dependence of $\mu_0H_{irr}$ can be fitted with $\mu_0H_{irr}(T) = \mu_0H_{irr}(0)(1-t)^{\beta}$, where $t = T/T_c$, which gives $\beta = 1.07$ for FeS and $\beta = 1.12$ for FeS$_{0.94}$Se$_{0.06}$. There are two primary mechanisms of core pinning from spatial variation of the Ginzburg-Landau (GL) coefficient $\alpha$ in type $\rm\uppercase\expandafter{\romannumeral2}$ superconductors, corresponding to the spatial variation in transition $T_c$ ($\delta$T$_c$ pinning), or to charge carrier mean free path $l$ near lattice defects ($\delta$$l$ pinning). In the case of $\delta$T$_c$ pinning, $J_c(t)$/$J_c(0)$ $\propto$ (1-$t^2$)$^{7/6}$(1+$t^2$)$^{5/6}$, whereas $J_c(t)$/$J_c(0)$ $\propto$ (1-$t^2$)$^{5/2}$(1+$t^2$)$^{-1/2}$ for $\delta$$l$ pinning, where $t$ = $T/T_c$.\cite{Griessen} As shown in Fig. 4(d), the reduced temperature dependence of reduced critical current density is nearly the same for FeS and FeS$_{0.94}$Se$_{0.06}$ and is between the theoretical curves of $\delta$T$_c$ pinning and $\delta$$l$ pinning. This suggests the presence of both microscopic mechanisms. Each contribution can be estimated by $J_{c, H=0}(t) = xJ_{c, H=0}^{\delta{T_c}} + (1-x)J_{c, H=0}^{\delta{l}}(t)$. The fitting gives $x$ = 0.15 for FeS, and $x$ = 0.17 for FeS$_{0.94}$Se$_{0.06}$, indicating $\delta$$l$ pinning plays major role in FeS and FeS$_{0.94}$Se$_{0.06}$. The pinning mechanism is different from K$_x$Fe$_{2-y}$Se$_2$, where $\delta$$T_c$ pinning is prevalent, and is similar with YBa$_2$Cu$_3$O$_7$ and NaFe$_{0.97}$Co$_{0.03}$As.\cite{LeiHC2,Griessen,Shabbir} Due to intrinsic nanoscale phase separation in K$_x$Fe$_{2-y}$Se$_2$\cite{Ryan,Wang,Liu,Ricci,Li} $\delta$$T_c$ could play a major role in pinning, in contrast to tetragonal iron sulfide even though FeS and FeSe$_{0.06}$ single crystals are obtained from K$_x$Fe$_{2-y}$(Se,S)$_2$ by de-intercalation. Moreover, because FeS is a typical type $\rm\uppercase\expandafter{\romannumeral2}$ superconductors, the multigap might have an effect on the fitting result. Nevertheless, the Dew-Hughes model still reveals that the $\Delta{l}$ pinning by Dew-Hughes model still gives an insight in the pinning mechanism.

\section{CONCLUSIONS}

In summary, we report the increase in the critical current density in tetragonal FeS single crystals by Se doping. The core normal surface-like pinning is present in the vortex dynamics. The pinning is dominated by the variation of charge carrier mean free path $l$ near lattice defects ($\delta$$l$ pinning). The critical current density is comparable to iron based superconductors with much higher superconducting transition temperature. This suggests that FeS-based superconducting structures with higher $T_{c}$'s could exhibit high performance, potentially attractive for low temperature high magnetic field applications.

\section*{Acknowledgements}
The work carried out at Brookhaven was primarily supported by the Center for Emergent Superconductivity, an Energy Frontier Research Center funded by the U.S. DOE, Office of Basic Energy Sciences (A.W and C.P) but also in part by the U.S. DOE under Contract No. DEAC02-98CH10886. This work has also been supported by Grant No. 171001 from the Serbian Ministry of Education and Science. Jianjun Tian was in part supported by a scholarship of Faculty Training Abroad Program of Henan University.

\end{document}